# Intrinsic electromagnetic damping in superconductor-ferromagnet proximity heterostructures

D. Seleznyov, Ya. Turkin, N. Pugach

*HSE University, 101000, Moscow, Russia*

L. Tao

*School of Physics, Harbin Institute of Technology, 150001, Harbin, China*

The study of the response of superconducting hybrid structures with magnetic materials to microwave irradiation is necessary for the development of effective superconducting spintronic devices. The role of the magnetic proximity effect (direct and inverse) on the electrical properties of hybrid structures is a pressing issue for its application. We theoretically study the electromagnetic impedance of a thin superconducting (S) film covering a ferromagnetic insulator (FI). An intrinsic damping of microwave irradiation is predicted because of the inverse proximity effect. The system of Usadel equations is solved numerically and self-consistently in the Nambu-Keldysh formalism with boundary conditions for strong spin polarization of the insulator. Based on the calculated Green's function, the features of the bilayer complex conductivity and impedance as a function of the field frequency have been discovered. It is shown how the ferromagnetic proximity effect leads to irremovable damping in the electromagnetic response of such heterostructures. The mechanism related to the formation of triplet Cooper pairs in an S layer at proximity of the FI interface is analyzed. Even gapless superconductivity has been found in a thin S film similar to a magnetic superconductor. The resulting intrinsic damping must be taken into account when designing superconducting devices for microwave applications.

For example, the proximity effect may be suppressed by protection of the S-FI interface with an additional thin insulating layer, transparent for irradiation.

## I. Introduction

The application of superconductors in nanoelectronic devices today is an extremely popular vector in the development of computer technology. This field is particularly intriguing due to the possibility of control of the macroscopic quantum phenomena required for constructing the quantum logic circuits[1-2]. Another prospective application of the superconductors lies in the field of spintronics. The proximity effects observed near the interface of a superconductor and adjacent magnetic material allow the generation and the manipulation of the spin currents. In such bilayers, the presence of a ferromagnet (F) provides additional spin asymmetry of Cooper pairs, resulting in the generation of triplet spin currents at magneto dynamics [3-5]. Manipulating these currents could potentially enable the development of effective cryogenic memory components and spintronic devices [6-13].

Rich theoretical possibilities to control the superconducting properties of the hybrid nanostructures with the external electromagnetic field, have stimulated experimental and theoretical research during last few years. [14-24]. It has been shown that the mutual influence between the superconductor and adjacent ferromagnetic insulator plays a crucial role in the coupling between the collective magnetization excitations through triplet superconducting correlations [18,19]. The pure electrodynamic proximity effect, which emerges from the induced Meissner currents inside the superconductor/ferromagnet structures with different directions of magnetization [20] gates the spin waves [21], and changes the magnon spectrum [22]. also, it has been predicted that the pronounced electromagnetic proximity effect and spin splitting field can cause the effective Dzyaloshinskii-Moriya interaction [23] and Andreev current enhancement [24]. Such electromagnetic proximity effect perturbs the self-magnetostatic field of the ferromagnetic insulator and may change the linear response. Here we consider only the electric response of thin enough superconducting film to examine the influence of the proximity effect with a ferromagnet. The magnetic response of a nonmagnetic (S) layer seems to be much less pronounced in the electromagnetic wave propagation. In addition, the Meissner response of a thin film may be neglected.

The research of the structures based on the contact of a ferromagnetic insulator and a superconductor [25] uncovers the potential achievement of the strong magnon-photon coupling inside the microwave resonators. Moreover, work [26] theoretically demonstrates that the interaction of the electromagnetic field with different superconducting structures unlocks the new types of pairing symmetries which opens the potential way for tuning the quantum properties of the materials.. Such a rich picture of the physical effects opens broad possibilities to design nanoelelctronic elements for the magnonic signal processing [22], quantum interferometry [27], neuromorphic circuits [28], and thermoelectric devices

[29] etc. Thus, the nonequilibrium physics of superconducting structures is a rapidly growing domain of condensed matter physics, which provides the obvious advantages for the design of the spintronic, magnonic, and quantum devices. In addition, hybrid superconductor structures play a crucial role in the investigation of the superconducting diode effect [30] for its practical use. This brief review is far from being full as the number of works in this field is growing everyday.

Superconducting spin triplet Cooper pairs can arise in superconductor-ferromagnetic insulator (S-FI) structures, as a consequence of the inverse proximity effect [31-32]. In this configuration, the band gap of the FI prevents the penetration of the superconducting condensate, causing a spin-dependent reflection at the S/FI interface, which leads to the conversion of singlet Cooper pairs into triplet ones. This inverse proximity effect is larger with ferromagnetic insulators than ferromagnetic metal, allowing a drainage of superconducting condensate, both singlet and triplet, into the metal (usual proximity effect) [33]. The inverse proximity effect exhibits itself as a suppression of the order parameter, the appearance of induced magnetization, and the associated zero bias peak (ZBP) formation in the density of states. It is crucial that the inverse proximity effect should also affect the electromagnetic response of S-FI structures. This aspect becomes significant during the design of metamaterials containing superconducting layers, because the prevalence of the imaginary part of conductivity in the superconducting phase allows the miniaturization of superconducting nanostructures without increasing the undesirable losses [34-35].

## II. Theoretical model

The early works [36-37] provide calculations of the microwave properties of S-F layers in a two-fluid model without taking into account the quantum nature of proximity effects. In this work, we use the quasiclassical theory of superconductivity as the basis for a theoretical model of the S-FI structure to calculate the bilayer complex conductivity $\sigma(\omega)$ and electromagnetic impedance. $Re(\sigma(\omega))$ defines the energy dissipation in the system and the related absorption of electromagnetic waves at low frequencies $\omega$ of the order of the superconducting spectrum gap $\Delta$, at whose the superconducting properties manifest themselves. At $\omega \gg \Delta$ superconductivity is destroyed by the irradiation. The physical behavior of the system is determined by the Green function $\hat{G} \equiv (\hat{G}^R \ \hat{G}^K \ 0 \ \hat{G}^A)$ in the Nambu-spin-Keldysh space [38], which includes the retarded $\hat{G}^R$, advanced $\hat{G}^A$ and Keldysh $\hat{G}^K$ elements. These components form 4 by 4 Green's function matrices in Nambu-spin space. If failing to consider nonequilibrium impacts in the dirty limit, the retarded Green's function must satisfy the Usadel differential equation of the following form:

$$iD\partial_x(\hat{G}^R(\varepsilon)\partial_x\hat{G}^R(\varepsilon)) = [\varepsilon\hat{\tau}_z + \hat{\Delta}(x), \hat{G}^R] \quad (1)$$

Here $D$ is diffusion coefficient, $\hat{\tau}_z$ – third Pauli matrix in spin space, $\hat{\Delta}$ – matrix of the order parameter $\Delta$ in diagonal form, and $x$ – spatial coordinate in the superconducting layer in the direction normal to the layer plane. Parameter $\varepsilon = \epsilon + i\eta$ is the sum of the electron kinetic energy $\epsilon$ and an inelastic energy dissipation constant $\eta$. In this approximation, the advanced $\hat{G}^A$ and Keldysh $\hat{G}^K$ components identified accordingly by relations $\hat{G}^A = -[\hat{\tau}_z\hat{G}^R\hat{\tau}_z]^\dagger$ and $\hat{G}^K = \hat{G}^R\hat{p} - \hat{p}\hat{G}^A$, where $\hat{p} = tanh(\varepsilon/2T)$. The last one defines the equilibrium distribution function allowing for the calculation of the linear response.

To solve the Usadel equation in the case of contact of a superconductor with a ferromagnetic insulator, it is necessary to use boundary conditions for a strong spin polarized insulator. The border with the external environment (air or vacuum) is described by the zero value of the derivative of the retarded Green's function. The boundary condition at the S-FI interface is given by the equality of the matrix currents through the interface, which is described by the formulas [31]:

$$\partial_x\hat{G}^R = 0, \qquad GL(\hat{G}^R\partial_x\hat{G}^R) = I(\varphi, \hat{G}^R) \quad (2)$$

Here, $G$ represents the overall conductivity of the material and $L$ denotes the superconductor layer thickness. The boundary conditions at the S-FI interface are contingent upon the spin mixing angle φ that specifies the acquired phase difference for the electrons with oppositely directed spins reflected from the FI boundary [31-32]. For computer calculations, it is customary to use the Riccati parameterization of $\hat{G}^R$, which automatically satisfies the normalization condition $\hat{G}^{R^2} = \hat{1}$ [38].

$$\hat{G}^R = \begin{pmatrix} N(\gamma, \tilde{\gamma}) & 0 \\ 0 & \tilde{N}(\gamma, \tilde{\gamma}) \end{pmatrix} \cdot \begin{pmatrix} 1 + \gamma\tilde{\gamma} & 2\gamma \\ 2\tilde{\gamma} & 1 + \tilde{\gamma}\gamma \end{pmatrix} \quad (3)$$

Where $N(\gamma, \tilde{\gamma}) = (1 - \gamma\tilde{\gamma})^{-1}, \tilde{N}(\gamma, \tilde{\gamma}) = (1 - \tilde{\gamma}\gamma)^{-1}$ are normalization matrices and $\gamma = \gamma(x, \varepsilon), \tilde{\gamma} = \tilde{\gamma}(x, \varepsilon)$ are Riccati parameters. The density of states is determined by the following formula $N(\varepsilon) = \frac{1}{2}Sp\{Re(N(\gamma, \tilde{\gamma})(1 + \gamma\tilde{\gamma}))\}$.

The Usadel equation and its boundary conditions can be represented in a parameterized form [35]:

$$D(\partial_z^2\gamma + 2(\partial_z\gamma)\tilde{\gamma}N(\partial_z\gamma)) = -i(V_{11}\gamma - \gamma V_{22} + \gamma V_{21}\gamma - V_{12})$$
$$D(\partial_z^2\tilde{\gamma} + 2(\partial_z\tilde{\gamma})\gamma\tilde{N}(\partial_z\tilde{\gamma})) = i(V_{22}\tilde{\gamma} - \tilde{\gamma}V_{11} + \tilde{\gamma}V_{12}\tilde{\gamma} - V_{21})$$

$$2GLN\partial_z\gamma = I_{12} - I_{11}$$
$$2GLN\partial_z\tilde{\gamma} = I_{21} - I_{22} \quad (4)$$

Here the components $V_{11}$, $V_{12}$, $V_{21}$, $V_{22}$ are determined from the following matrix 2x2:

$$V = \varepsilon \hat{\tau}_z + \hat{\Delta}(z) \quad (5)$$

Equations (4) must be supplemented with a self-consistency condition for the s-wave order parameter, which contains only singlet components of the anomalous Green function:

$$\frac{\Delta(z)}{\Delta_0} = \frac{1}{2} n_0 \lambda \int_0^{sinh(1/n_0\lambda)} d\left(\frac{\epsilon}{\Delta_0}\right) [(N\gamma)_{12} - (N\gamma)_{21} - \left(\widetilde{N}\widetilde{\gamma}\right)_{12}^* + \left(\widetilde{N}\widetilde{\gamma}\right)_{21}^*] \cdot tanh\left(\frac{\pi}{2e^c} \frac{\frac{\epsilon}{\Delta_0}}{\frac{T}{T_c}}\right) \quad (6)$$

Where $c$ is the Euler-Mascheroni constant, $n_0$ is the density of states at the Fermi level in normal state, $\lambda$ is the pairing constant of the BCS theory, and $\Delta_0$ is the order parameter of a bulk superconductor at low temperature, for which the calculations are carried out.

The expression of the superconductor impedance can be calculated in the nonequilibrium Keldysh formalism [39]. It is related to conductivity through the relationship $Z(\omega) = \sqrt{\frac{\omega}{i\sigma(\omega)}}$.

Let us assume the normal incidence of an electromagnetic wave on a superconducting film placed on a FI layer. No additional non-uniform change of the phase and the amplitude of the order parameter is produced inside the condensate [40-41]. Working in the Coulomb gauge and using the definition of the electric current, the impedance and the response kernel of the dirty superconductor can be expressed through the unperturbed retarded and advanced Green functions calculated from the stationary task, as it has been done in the work [42]. The distribution function enters the expression through the normalization condition for the stationary task. In the time-domain, we can write the response of the superconductor to the plane electromagnetic wave with in-plane electric field as:

$$j = -i\frac{\sigma_0}{4} Sp\{\hat{\tau}_z \hat{g}^K \circ \hat{A} \circ \hat{g}^A + \hat{\tau}_z \hat{g}^R \circ \hat{A} \circ \hat{g}^K\}_{t_1 = t_2 = t} \quad (7)$$

Where the $\circ$ is the time-convolution operator, $\sigma_0 = 2e^2 n_0 D$ is the conductance of the metal layer in normal state, $e$ – is the electron charge, $\hat{g}^A, \hat{g}^R, \hat{g}^K$ are the time-dependent corresponding components of the Green function $\hat{G}$. Matrix of the vector potential $\hat{A}$ is defined in terms of $\hat{\tau}_3$ matrix:

$$\hat{A} = A(t_1)\hat{\tau}_3 \delta(t_1 - t_2) \quad (8)$$

Let the electric field changes in time as a function of the frequency of the microwave signal $\omega$ as $E_z = E_0 sin(\omega t)$, where $E_0$ - is the amplitude of the electric part of the plane electromagnetic wave. After the Fourier transform of formula (7), we get the following expression for the conductivity through $\hat{G}^R$, $\hat{G}^A$ and $\hat{G}^K$:

$$\sigma(\omega) = -\frac{\sigma_0}{16\omega\hbar} \int d\varepsilon Sp \left\{\hat{\tau}_z \hat{G}^K \left(\varepsilon - \frac{\hbar\omega}{2}\right) \hat{\tau}_z \hat{G}^A \left(\varepsilon + \frac{\hbar\omega}{2}\right) + \hat{\tau}_z \hat{G}^R \left(\varepsilon - \frac{\hbar\omega}{2}\right) \hat{\tau}_z \hat{G}^K \left(\varepsilon + \frac{\hbar\omega}{2}\right)\right\} \quad (9)$$

The set of equations (4) is numerically solved together with the self-consistent equation (6) via an iteration procedure. Substituting the numerically found retarded Green's function $\hat{G}^R$ into the resulting expression (9), we can calculate the frequency dependences of the complex conductivity components of the S-FI bilayer.

### III. Results

For our calculations, we choose Aluminum as a superconductor with coherence length $\xi \approx 100$ $nm$ in the dirty case. The model was characterized by the following parameters: the areal density $-\frac{N}{A} = 2.48 \cdot 10^{19}$ $\mu m^{-2}$, the inelastic scattering parameter was $\eta = 0.02$. The system was operated at a temperature close to absolute zero, while the critical temperature for Al was set at $T_c = 1.2$ K. In this work, the length of the S layer was chosen to be 120 nm=1.2 $\xi$ to enhance the influence of the inverse proximity effect in the S-FI bilayer, which has a scale of the coherence length. At the same time, the S film should not be too thin to keep superconductivity suppressed by the inverse proximity effect. Realistic values of the spin mixing angle serving as the S-FI interface parameter were estimated to be small enough from available experiments, φ<<1 [32].

Both the density of states and the superconducting order parameter depend on the distance from the FI interface at the proximity case. The order parameter is suppressed by the inverse proximity effect due to singlet-to triplet conversion close to the FI interface, where the triplet features are also the most pronounced in DoS. The Green function, as well as the complex conductivity, depends on this distance. For the analysis of the integral characteristics, the total electromagnetic response and total or averaged over the film thickness values were calculated. FIG.1a. demonstrates the non-monotonic behavior of the average DoS over the superconducting layer depending on the degree of conversion of singlet to triplet Cooper pairs given by the parameter $\varphi$. In the absence of the inverse proximity effect, with $\varphi$ equal to zero, DoS corresponds to a bulk superconductor from the BCS theory. For a small $\varphi = 0.004$, a characteristic zero bias peak is formed at the Fermi energy, signaling the presence of triplet superconductivity in the structure.

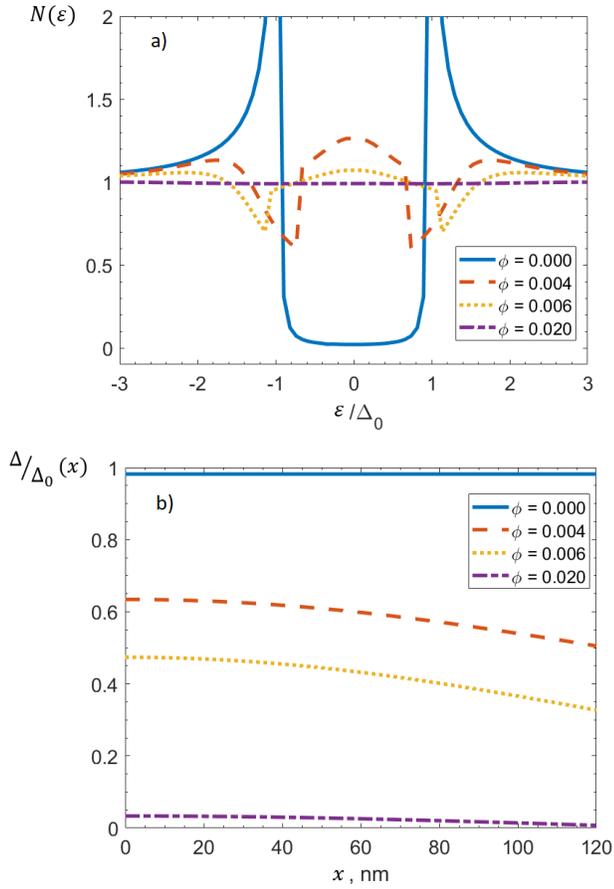

FIG.1. (a) The average DoS in the S layer for different values of spin-mixing angle φ. (b) The distribution of the order parameter in the S layer. The FI layer is located at the right side at $x=120$ nm.

This peak is a consequence of the specific time-reversal antisymmetry of triplet superconducting components, which are symmetric with respect to spin permutation [31, 43]. Further, as the spin-mixing angle increases, a suppression in superconductivity occurs, leading to a decrease in the height of the ZBP and an increase in its width. At higher φ > 0.02, superconductivity in the thin S film is totally suppressed. Similar manifestations of non-monotonicity in the features of DoS and induced magnetization depending on the parameters were studied in the work [45]. Note that DoS does not reach zero everywhere in the range of the gap of the S film. It means the absence of the gap in the electron spectrum. However, the superconducting order parameter is still nonzero. It is only partially suppressed at intermediate values of φ (FIG.2b). The superconductivity is still present and regular, not just in terms of superconducting fluctuations. So, the inverse proximity effect creates triplet superconducting correlations, caused in the induced magnetization, and penetrating the entirety of the thin S film. They also result in gapless superconductivity like in ferromagnetic superconductors. Note that here the nonzero order parameter is only formed by singlet superconducting correlations. These magnetic properties of a simple singlet superconducting film, induced by the inverse proximity effect with the FI, should lead to peculiarities in the complex conductivity and electromagnetic response. In fact, electromagnetic waves propagate in this artificial "induced triplet superconductor".

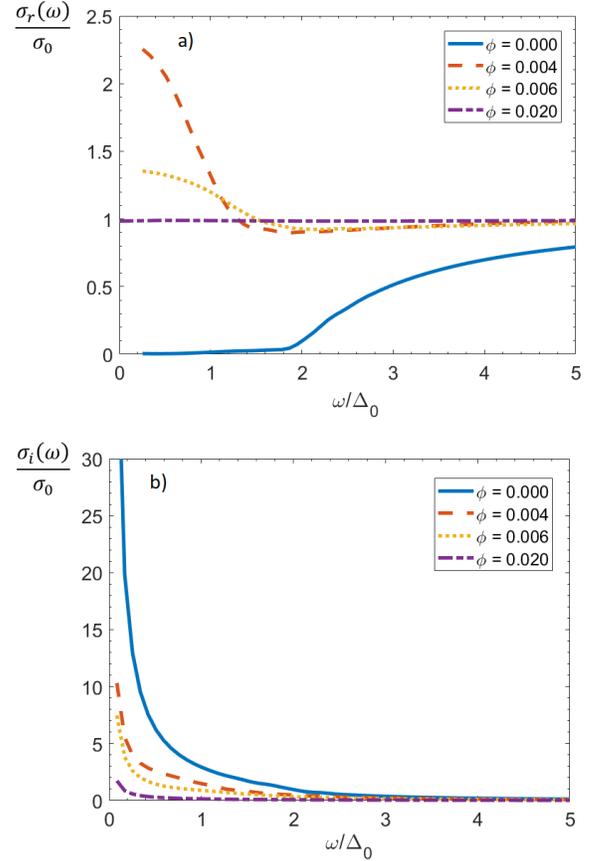

FIG.2. The real (a) and imaginary (b) part of average complex conductivity $\sigma(\omega)$ (in $\sigma_0$ units) as a function of the frequency of the microwave signal $\omega$ for different values of spin-mixing angle φ.

Figure 2 shows the plots of averaged complex conductivity $\sigma(\omega) = \sigma_r(\omega) - i\sigma_i(\omega)$ at different spin mixing angles. The frequency $\omega$ is specified in units of the superconducting gap of the bulk Aluminum at zero temperature $\Delta_0$. For $\varphi = 0$, the frequency response of the superconductor relates to the behavior of a bulk superconductor in an oscillating field. At $\omega$ equal to zero, conductivity is determined by the delta function $\delta(\omega)$. A characteristic growth in the real part of the conductivity is noticeable at a frequency of $\omega_\Delta = 2\Delta_0$. The characteristic frequency here is $\omega_\Delta = 88$ GHz, which corresponds to a wavelength of $\lambda_\Delta = 3.4$ mm for Aluminum. In the interval $\omega > \omega_\Delta$, superconductivity is destroyed by a high-frequency field. The imaginary part of the high-frequency conductivity behaves as a function of $\sigma_i \sim \frac{1}{\omega}$ and becomes equal to zero at $\omega > \omega_\Delta$.

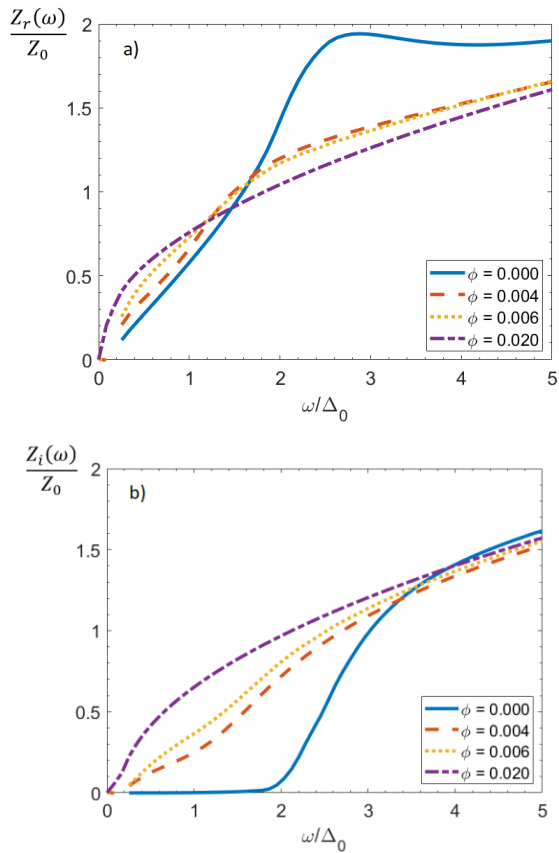

FIG.3. The real (a) and imaginary (b) part of average complex impedance $Z(\omega)$ (in $Z_0$ units) as a function of the frequency of the microwave signal $\omega$ for different values of spin-mixing angle φ.

Figure 3 shows the plots of averaged complex impedance $Z(\omega) = Z_r(\omega) - iZ_i(\omega)$ at different spin mixing angles. The $Z_r$ component describes the surface resistance of the structure, while $Z_i$ is the surface reactance, and $Z_0 = 1/\sigma_0$. The $Z_r$ is associated with the loss of electromagnetic wave energy reflected from a superconductor and the scattering mechanism of normal carriers. The imaginary part $Z_i$ characterizes the response of superconducting carriers. At low frequencies and $\varphi = 0$, the reactance of a superconductor is very close to zero due to its ability to efficiently conduct electric current without loss and with minimal interactions with magnetic fields. However, with increasing frequency, there is a sharp increase in reactance, and then for frequencies greater than $\omega_\Delta$, there is a smooth increase in reactance with increasing frequency, as in normal metal. For non-zero spin mixing angles, a smooth increase in reactance also occurs in the region $\omega < \omega_\Delta$. Interesting to note that the surface resistance $Z_r$ of ferromagnetically proximized S film ($\varphi = 0.004, 0.006$) exceeds both the surface resistance of a bulk superconductor ($\varphi = 0$) and the metal in normal state ($\varphi = 0.02$) in some frequency range at $\omega \sim \Delta_0$. The surface reactance $Z_r$ has only an inflection point in this frequency range, but becomes a little less than one of pure superconducting or normal state at $\omega > 3\Delta_0$.

The influence of a ferromagnetic insulator on a superconductor is expressed in the suppression of superconductivity and generation of spin triplets, which in turn affects the impedance and complex conductivity of this structure in a high-frequency field. At $\varphi = 0.02$ superconductivity is fully suppressed, and the behavior corresponds to the response of a normal layer. As the spin mixing angle increases, a gradual suppression of the imaginary part of the conductivity $\sigma_i$ is observed, while the conductance's real component $\sigma_r$, related to dissipation, exhibits particularly intriguing behavior. At frequencies higher than characteristic jump $\omega_\Delta$, component $\sigma_r$ corresponds to conductivity in the normal state. It is surprising that for frequencies lower than $\omega_\Delta$, the conductivity $\sigma_r$ exceeds the value $\sigma_0$ few times at intermediate spin mixing angle φ ($\varphi = 0.004, 0.006$). This increase in conductivity $\sigma_r$ can be explained by the delta function blurring and the presence of the electrons with energy less than the gap energy $\Delta$ in the system. The peak of the real part of the conductivity shows similar behavior as the ZBP on the density of states (FIG.1.). The zero bias peak contains several electron states under the gap. Excitations in these states turn the dissipation of electromagnetic waves energy in the S film at low frequencies, much lower than the superconducting spectrum gap. The gap may even be absent in a singlet superconductor in the proximity with a ferromagnetic insulator. Similar conductivity enhancement at frequencies lower than $\omega_\Delta$ has been found theoretically and experimentally in p-wave superconductors (strontium ruthenate), where ZBP appears due to triplet pairing [44, 45].

Previously, our analytical calculation demonstrated a non-monotonic dependence of magnetization on the spin mixing angle, which was further confirmed by numerical calculations of the features of the density of state [32, 46]. The complex conductivity of the S-FI structure also exhibits non-monotonic patterns in this region, indicating fundamental physical characteristics of such properties that may be especially noticeable in thin S films with a thickness of the order of the coherence length. In heterostructures with thicker S layers, the electromagnetic losses would locate in the $\xi$ -proximity of the magnetic interface. If the magnetic layer is metallic, the effect would be maintained, but with a different specific value, because it is derived from the general features of the magnetic proximity effect like singlet-to triplet conversion.

In summary, the system of Usadel equations is solved numerically and self-consistently in the Nambu-Keldysh formalism with boundary conditions for strong spin polarization of the insulator. Based on the calculated Green's function, the features of the bilayer complex conductivity and surface impedance as functions of the field frequency were discovered. It is shown that the ferromagnetic proximity effect leads to irremovable damping in the electromagnetic response of such heterostructures. Due to the inverse proximity effect, the ferromagnetic insulator (FI) affects the density of states (DoS) of the superconducting layer (S), caused by the conversion of singlet Cooper pairs into triplet ones. It manifests itself in the formation of a characteristic zero bias peak for energies lower than the gap energy. No thick superconducting film occurs under the conditions of gapless superconductivity, similar to magnetic superconductors. It should cause the appearance of similar features of the electromagnetic impedance like an intrinsic damping of irradiation.

Singlet-to triplet conversion due to the inverse proximity effect with a ferromagnetic insulator makes the usual s-wave superconducting film similar to a triplet magnetic superconductor in many respects: the presence of induced magnetization, zero bias peak in DoS, and electromagnetic wave absorption. This implies that incorporating S-FI structures in superconducting microwave devices will lead to extra losses caused by absorbing electromagnetic irradiation by under-gap excitations. These features are intrinsic for such structures and are the consequences of s-wave superconductivity and the inverse proximity effect. To exclude such losses, the interface with a ferromagnet (insulating or metallic) may be protected by an extra thin (nonmagnetic) insulating layer, which suppresses the proximity effect, but is transparent for microwaves.

**Acknowledgements**

The calculations and the text have been supported by Russian Science Foundation project 23-72-00018 "Study of non-equilibrium and boundary phenomena in superconducting hybrid nanostructures". The participation of L. Tao in the results discussion and the manuscript preparation is supported by the Fundamental Research Funds for the Central Universities (FRFCU5710053421) and the National Natural Science Foundation of China (Grant No. 12274102).

## References

[1] S. Razmkhah, Beyond-CMOS: State of the Art and Trends, 295–391 (2023).
[2] Klenov et al., J. Phys.: Conf. Ser. 97, 012037 (2008).
[3] T. Tokuyasu, J. A. Sauls, and D. Rainer, Phys. Rev. B 38, 8823–8833 (1988).
[4] E. A. Demler, G. B. Arnold, and M. R. Beasley, Phys. Rev. B 55, 15174–15182 (1997).
[5] Y. V. Turkin and N. Pugach, Beilstein J. Nanotechnol. 14, 233-239 (2023).
[6] J. Linder and J. W. A. Robinson, Nat. Phys. 11, 307–315 (2015).
[7] M. Eschrig, Rep. Prog. Phys. 78, 104501 (2015).
[8] V. Vedyayev, N. V. Ryzhanova, and N. G. Pugach, J. Magn. Magn. Mater. 305, 53-56 (2006).
[9] V. Gordeeva et al., Sci. Rep. 10, 21961 (2020).
[10] F. S. Bergeret, M. Silaev, P. Virtanen, and T. T. Heikkilä, Rev. Mod. Phys. 90, 041001 (2018).
[11] I. A. Golovchanskiy, N. N. Abramov, V. S. Stolyarov, A. A. Golubov, V. V. Ryazanov, and A. V. Ustinov, J. Appl. Phys. 127, 093903 (2020).
[12] I. A. Golovchanskiy, V. V. Ryazanov, and V. S. Stolyarov, Phys. Rev. Applied 20, L021001 (2023).
[13] I. A. Golovchanskiy, N. N. Abramov, O. V. Emelyanova, I. V. Shchetinin, V. V. Ryazanov, A. A. Golubov, and V. S. Stolyarov, Phys. Rev. Applied 19, 034025 (2023).
[14] J. Linder, M. Amundsen, and J. A. Ouassou, Sci. Rep. 6, 38739 (2016).
[15] Haxell et al., Nat. Commun. 14, 6798 (2023).
[16] B. Li, N. Roschewsky, B. A. Assaf, M. Eich, M. Epstein-Martin, D. Heiman, M. Munzenberg, J. S. Moodera, Phys. Rev. Lett. 110, 097001 (2013).
[17] S. Banerjee and M. S. Scheurer, Phys. Rev. Lett. 132, 046003 (2024).
[18] N. A. Gusev, D. I. Dgheparov, N. G. Pugach, and V. I. Belotelov, Appl. Phys. Lett. 118(23):232601 (2021).
[19] V. Bobkova, A. M. Bobkov, A. Kamra, and W. Belzig, Commun. Mater. 3, 95 (2022).
[20] S. Mironov, A. S. Mel'Nikov, and A. Buzdin, Appl. Phys. Lett. 113, 023605 (2018).
[21] M. Borst et al., Science 382, 430 (2023).
[22] Dobrovolskiy et al., Nature Physics 15, 477 (2019).
[23] M. A. Kuznetsov and A. A. Fraerman, Physical Review B 105, 214401 (2022).
[24] A. Ozaeta, A. S. Vasenko, F.W. J. Hekking, F. S. Bergeret, Physical Review B 86, 060509(R) (2012).
[25] L. McKenzie-Sell, J. Xie, C. M. Lee, J. W. A. Robinson, C. Ciccarelli, J.A. Haigh, Phys. Rev. B 99, 140414(R) (2019).
[26] J. Cayao, C. Triola, A. M. Black-Schaffer, Physical Review B, 103(10), 104505. (2021).
[27] Giazotto, F., Heikkilä, T. T., and Bergeret, F. S., Physical Review Letters 114, no. 6, 067001 (2015).
[28] M. Schneider et al., Science Advances 4, no. 1, e1701329 (2018).
[29] A. S. Vasenko et al., J. Low Temp. Phys. 154, 221 (2009).
[30] M. Nadeem et al., Nat. Rev. Phys. 5, 558 (2023).
[31] J. A. Ouassou et al., Sci. Rep. 7, 1932 (2017).
[32] V. O. Yagovtsev et al., Supercond. Sci. Technol. 34, 025003 (2021).
[33] S. M. Anlage, IEEE J. Microwaves 1, 389-402 (2021).
[34] S. M. Anlage, J. Opt. 13, 024001 (2010).
[35] J. Wu and Y. L. Chen, Prog. Electromagn. Res. 111, 433-445 (2011).
[36] T. Nurgaliev, Physica C 468, 912-919 (2008).
[37] V. Chandrasekhar, Springer-Verlag, V 2 (2004).
[38] S. H. Jacobsen, J. A. Ouassou, J. A. Linder, Phys. Rev. B 92, 024510 (2015).
[39] P. I. Arseev, Phys.-Uspekhi 58, 1159 (2015).

[40] P. I. Arseev, S. O. Loiko, and N. K. Fedorov, Phys.-Uspekhi 49, 1 (2006).
[41] A. A. Radkevich, A. G. Semenov, Phys. Rev. B 106, 094505 (2022).
[42] Y. V. Fominov, M. Houzet, and L. I. Glazman, Phys. Rev. B 84, 224517 (2011).
[43] Y. Tanaka, A. A. Golubov, Phys. Rev. Lett. 98, 037003 (2007).
[44] Y. Asano, A. A. Golubov, Y. V. Fominov, Y. Tanaka, Phys. Rev. Lett. 107, 087001 (2011).
[45] S. V. Bakurskiy et al., Phys. Rev. B 98, 134508 (2018).
[46] D. V. Seleznyov et al., J. Magn. Magn. Mater. 171645 (2023).